\documentclass[epj,referee]{svjour}
\usepackage{amssymb}
\usepackage{axodraw}
\usepackage{rotate}
\usepackage{epsfig}
\usepackage{cite}
\usepackage{graphicx}
\usepackage{floatflt}

\begin{document}
\setcounter{page}{0}
\thispagestyle{empty}
\newcommand{\sanc}{{\tt SANC} }
\newcommand{\bq}{\begin{equation}}
\newcommand{\eq}{\end{equation}}
\newcommand{\bqa}{\begin{eqnarray}}
\newcommand{\eqa}{\end{eqnarray}}
\newcommand{\nll}{\nonumber\\}
\newcommand{\sss}[1]{\scriptscriptstyle{#1}}
\def\Ps {P^2}
\def\Qs {Q^2}
\def\Ts {T^2}
\def\Us {U^2}
\def\mw {M_{\sss W}}
\def\mz {M_{\sss Z}}
\def\mh {M_{\sss H}}
\def\stw{s_{\sss W}}
\def\ctw{c_{\sss W}}
\newcommand{\hsm}{\hspace*{-1mm}}
\def\gw {\Gamma_{\sss W}}
\newcommand{\GeV}{\unskip\,\mathrm{GeV}}
\def\href#1#2{#2}
\newcommand{\mtpll}{m_t^2}
\newcommand{\mbtll}{m_b^2}
\newcommand{\mtp}{m_t}
\newcommand{\mbt}{m_b}
\newcommand{\mupll}{m_{u}^2}
\newcommand{\mdnll}{m_{d}^2}
\newcommand{\mup}{m_{u}}
\newcommand{\mdn}{m_{d}}
\newcommand{\mwll}{M_{\sss W}^2}
\newcommand{\mws}{\hat{M}^2_{\sss W}}
\newcommand{\qup}{Q_u}
\newcommand{\qdn}{Q_d}
\newcommand{\qtp}{Q_t}
\newcommand{\qbt}{Q_b}
\newcommand{\scff}[1]{C_{#1}}
\newcommand{\Lnla}{\ln_\lambda}
\newcommand{\tmW}{{\widetilde{M}}_{\sss W}}
\newcommand{\Litwo}{\mbox{${\rm{Li}}_{2}$}}
\newcommand{\atan}{{\rm{atan}}}
\newcommand{\MeV}{\unskip\,\mathrm{MeV}}
\newcommand{\ds}{\displaystyle}

\title{SANCnews: top decays in QCD and EW sectors \thanks{
This work is partly supported by RFFI grant $N^{o}$07-02-00932-a} 
}

\author{
\and D. Bardin\inst{1} 
\and S. Bondarenko\inst{2} 
\and P. Christova\inst{1} 
\and L. Kalinovskaya\inst{1} 
\and V. Kolesnikov\inst{1}
\and W. von Schlippe\inst{3}
}

\institute
{Dzhelepov Laboratory of Nuclear Problems, 
JINR,\ Dubna, \ 141980 \ \  Russia
\and
Bogoliubov Laboratory of Theoretical Physics, 
JINR,\ Dubna, \ 141980 \ \  Russia
\and
PNPI, St. Petersburg, 188300 \ \ Russia
}

\abstract{In this paper we present the results of the implementation
of the decay $t \to b f_1\bar{f}'_1$ into the \sanc system ($f_1$ is a massless fermion). 
The new aspect of the work is the combination of QCD and EW corrections.
All calculations are done at the one-loop level in the Standard Model.
We give a detailed account of the new procedure --- the forming of a class of $J_{AW,WA}$ functions.
These functions are related to the procedure of extraction of infra-red and mass-shell singular divergences. 
The emphasis of this paper is on the presentation of numerical results for various approaches:
complete one-loop calculations and different versions of pole approximations.
}

\maketitle

\section{Introduction}
In this paper we review the state-of-the-art of the implementation of NLO QCD and electroweak (EW)
radiative corrections (RC) to the charge current decays
\bqa
F \to f f_1\bar{f}'_1(\gamma,g)
\label{Decaytlff}
\eqa
(where $F$ and $f$ denote massive fermions and $f_1$ and $f'_1$ denotes massless fermions)
within the framework of the SANC system~\cite{Andonov:2004hi},~\cite{Bardin:2005dp}.\\
\indent
This paper is a continuation of our previous one~\cite{Arbuzov:2007ke}, devoted to the EWRC
to $t\to b l^+\nu_l$ decay. Here we extend it in two directions: addition of quark channels
e.g. $t\to b u\bar{d}$ {\em etc} and of the NLO QCD corrections, see also Ref.~\cite{Sadykov:2006pp}
and references therein.
The implementation of QCD corrections into \sanc for some 3- and 4-leg processes is presented
in Ref.~\cite{Andonov:2006un}.

Recall that in SANC we always calculate any one-loop process amplitude as annihilation into vacuum with all
4-mo\-menta incoming. Therefore, the derived form factors for the amplitude of the process $t b\bar{u}\bar{d}\to 0$
after an appropriate permutation of their arguments may be used for the description of
NLO corrections of the single top production processes, e.g. $s$-channel $ud\to tb$, and $t$-channel $ub\to dt$.

The QCD tree for the $t\to bf_1\bar{f}'_1$ processes is shown in Fig.~\ref{Processes}:

\vspace*{-3mm}
\begin{figure}[!h]
\hspace*{5mm}
\includegraphics[width=7cm]{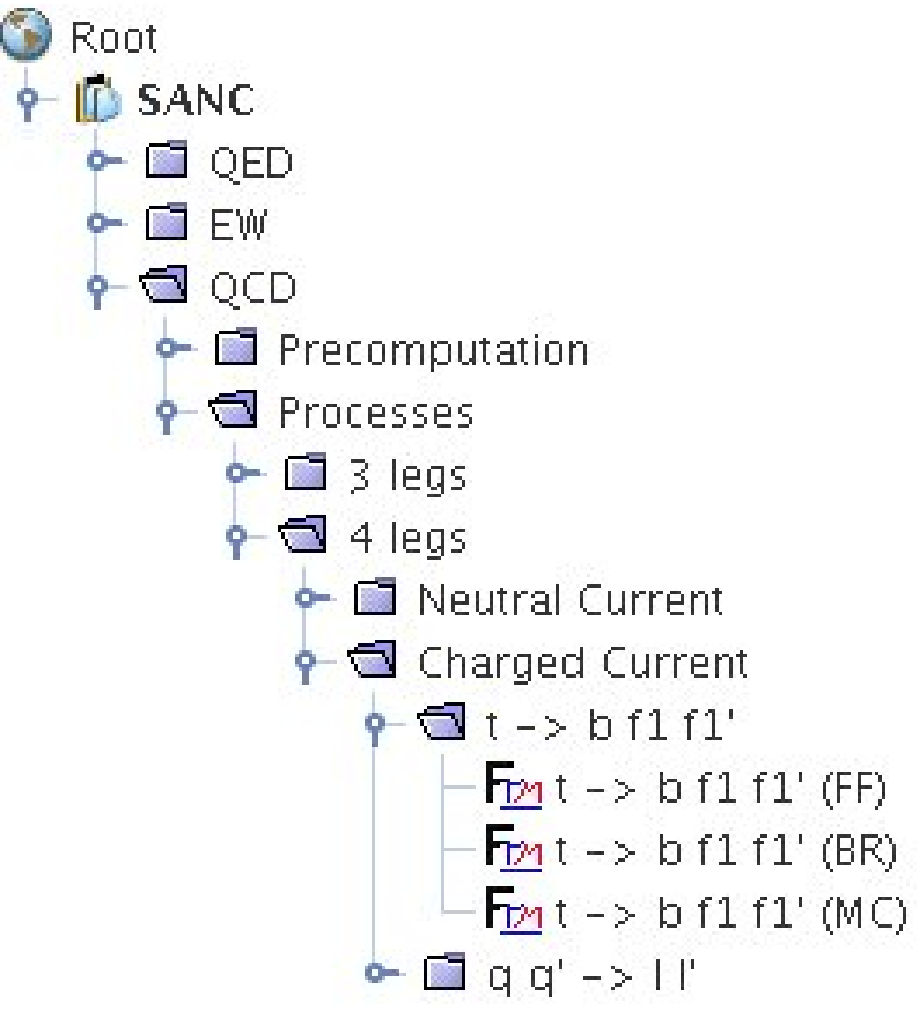}
\vspace*{-3mm}
\caption[QCD node: $t\to bf_1\bar{f}'_1$]
        {QCD node: $t\to bf_1\bar{f}'_1$}
\label{Processes}
\end{figure}
\vspace*{-6mm}

A similar tree was already shown in the previous paper~\cite{Arbuzov:2007ke} for the EW branch.
Nowadays, within \sanc we follow the strategy to present both EW and QCD NLO RC simultaneously, realizing
them as the SSFM (Standard SANC (FORM/FORTRAN) Modules).
We use FORM version 3.1~\cite{Vermaseren:2000nd}.
The modules are united into two packages (CC and NC).
The concept of modules is decribed in Ref.~\cite{Andonov:2008ga}, ibidem the
previous versions 1.20, see also~\cite{ACATVK}.

The packages are reachable for users from our project homepages~\cite{homepagesSANC}.
Both EW and QCD RC modules of these processes $t \to b f_1\bar{f}'_1$
will be put into version 1.30 of the CC package.

 A first attempt to combine QCD and EW corrections within the SANC project was done for DY CC processes
and presented in talks at the ATLAS MC Working Group~\cite{Kolesnikov:2006talk}
and later on in the preprint~\cite{Andonov:2009nn}.

 This paper is devoted to the complete NLO QCD and EW radiative cor\-rec\-tions (EWRC) to the 4-leg top quark
decays $t\to b f_1\bar{f}'_1(\gamma,g)$.
 We  discuss also how the \sanc results of complete one-loop calculations compare with the results
of various approximate cascade approaches.

These exercises are necessary in order to make the right choice in the future:
how we would sew together NLO 4-leg and 3-leg building blocks, available in \sanc~\cite{Andonov:2004hi}.
For example 4-leg and 3-leg blocks in the description of a cascade of the type
$f_1\bar{f_1}\to HZ; Z\to \mu^+\mu^-$~\cite{Sapronov:2009prepar}
or two 4-leg blocks in $ud\to bt; t\to bl\nu$ {\em etc}~\cite{BBCKK:2009prep}.

This paper is organized as follows.
In section~\ref{OneloopEWRC} we review the complete calculations as adopted
within the \sanc framework. The standard narrow width cascade approach, that with a complex $W$ boson mass,
and the cascade in the pole approximation with a finite $W$ width are presented in section~\ref{Cascade}.
Numerical results are collected in section~\ref{Numerics}.
In section~\ref{Concl} we present our conclusions.

\section{Complete EWRC \label{OneloopEWRC}}
\subsection{The separation of QED corrections \label{QEDseparation}}
The complete one-loop EW corrections for $t(p_2) \to b(p_1) + u(p_3) + \bar{d}(p_4)$ decay
are calculated by the \sanc system as described in section 2.5 of Ref.~\cite{Andonov:2004hi}.
The covariant amplitudes ${\cal A}$ and helicity amplitudes ${\cal H}_{ijkl}$
are given by Eqs.~(43)--(46) with $D_\mu=-(p_1+p_2)_\mu$ and Eqs.~(47)--(50), respectively.
They are expressed in terms of four scalar
form factors: ${\cal F}_{\sss LL},\,{\cal F}_{\sss RL},\,{\cal F}_{\sss LD},\,{\cal F}_{\sss RD}$.
It is useful to extract the QED part from the complete EW amplitude.
Only the $LL$ form factor contains both QED and weak contributions:
\bq
{\cal F}_{\sss LL}=1+\frac{e^2}{16\pi^2}{\widetilde{\cal F}}^{\rm QED}_{\sss LL}
                   +\frac{g^2}{16\pi^2}{\widetilde{\cal F}}^{\rm weak}_{\sss LL}\,.
\eq
The other three form factors contain only weak parts.
There exists no gauge invariant separation of the QED part
from the entire $LL$ form factor. We choose it in the simplest and most natural form:
\bqa
&&{\widetilde{\cal F}}^{\rm QED}_{\sss LL} = \nll
&&2\Bigl[-\qup\qdn \Qs
   \scff{0}( - \mup^2, - \mdn^2,\Qs;\mup,\lambda,\mdn)\nll
&-&\qup \qtp  ( \Ts +\mtp^2 )
   \scff{0}( - \mup^2, - \mtp^2,\Ts;\mup,\lambda,\mtp)\nll
&+&\qup \qbt \Us
   \scff{0}( - \mup^2, - \mbt^2,\Us;\mup,\lambda,\mbt)\nll
&+&\qdn \qtp  ( \Us + \mtp^2 )
   \scff{0}( - \mdn^2, - \mtp^2,\Us;\mdn,\lambda,\mtp)\nll
&-&\qdn \qbt \Ts
   \scff{0}( - \mdn^2, - \mbt^2,\Ts;\mdn,\lambda,\mbt)\nll
&-&\qtp \qbt  ( \Qs + \mtp^2 )
   \scff{0}( - \mtp^2, - \mbt^2,\Qs;\mtp,\lambda,\mbt)
\Bigr]\nll
&-&\frac{3}{2}\Bigl[
        \qup^2 a_0^f(\mup)
       +\qdn^2 a_0^f(\mdn)
       +\qtp^2 a_0^f(\mtp)
\nll
&+&
        \qbt^2 a_0^f(\mbt)
\Bigr]       +\qup^2 \Lnla(\mup^2)
       +\qdn^2 \Lnla(\mdn^2)
\nll
&+&
        \qtp^2 \Lnla(\mtp^2)
       +\qbt^2 \Lnla(\mbt^2)\,,
\label{QED_FF}
\eqa
with $\scff{0}$ being the standard Passarino--Veltman
function~\cite{Passarino:1978jh},~\cite{Bardin:1999ak} and
\bq
a_0^f(m)\,=\,\ln\left(\frac{m^2}{\mu^2}\right)-1, \quad \Lnla(m^2)\,=\,\ln\left(\frac{m^2}{\lambda^2}\right),
\eq
where $\mu$ is the t'Hooft scale and $\lambda$ is a photon mass. The natural choice is $\mu=\mw$.
Furthermore, in Eq.~(\ref{QED_FF}) we use the standard \sanc definitions:
$Q_f=2I^{3}_{f}$ with $I^{3}_{f}$ being the weak isospin and
\bqa
Q^2&=&(p_1+p_2)^2,
\nll
T^2&=&(p_2+p_3)^2,
\nll
U^2&=&(p_2+p_4)^2,
\eqa
with momenta $p_{i}$ being defined in Fig.~\ref{WAbox}.

The  form factor ${\widetilde{\cal F}}^{\rm QED}_{\sss LL}$, as defined by
Eq.~(\ref{QED_FF}), contains all IR divergences
in four $\Lnla(m^2)$ functions, one for each photon emission from an external line,
and in six $C_0$ functions, one for each photon radiation interference term.
Moreover, all logarithmic mass singularities should be concentrated in the QED
part and all weak contributions must not
contain logarithmic mass singularities even at the amplitude level, having nothing to do with the KLN theorem.
Furthermore, the gauge non-invariance of the QED/weak separation is made manifest by the presence of
the t'Hooft scale.
 We prefer to keep terms with $a_0^f(m)$ in the QED contribution since they are mass singular.

\subsection{Auxiliary functions $J_{AW,WA}$ \label{JAWs}}
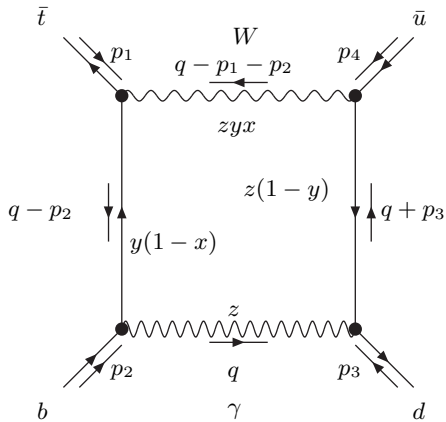
\begin{figure}[!h]
\[
\begin{picture}(132,132)(-20,0)

  \Text(-32,137)[lb]{$\bar t$}
  \Text(-32,-12)[lb]{$b$}
  \Text(110,137)[lb]{$\bar u$}
  \Text(110,-12)[lb]{$d$}

\Text(40,27)[lb]{$ z      $}
\Text(36,95)[lb]{$ zyx    $}
\Text(46,70)[lb]{$ z(1-y) $}
\Text(3,50 )[lb]{$ y(1-x) $}

\Text(42,130)[lb]{$W$}
\Text(20,118)[lb]{$q-p_1-p_2$}
\Text(40,2)[lb]{$q$}
\Text(40,-12)[lb]{$\gamma$}

  \ArrowLine(55,115)(33,115)
  \ArrowLine(33,17)(55,17)

  \ArrowLine(88,110)(88,22)
  \Vertex(88,22){2.5}
  \ArrowLine(88,22)(110,0)
  \Photon(88,110)(0,110){2}{10}

  \Vertex(0,110){2.5}
  \ArrowLine(0,110)(-22,132)
  \ArrowLine(0,22)(0,110)

 \Vertex(0,22){2.5}
 \ArrowLine(-22,0)(0,22)
 \Photon(0,22)(88,22){3}{15}

 \ArrowLine(104,132)(88,116)
 \Vertex(88,110){2.5}
 \ArrowLine(110,132)(88,110)
 \ArrowLine(104,0)(88,16)

 \ArrowLine(-16,0)(0,16)
 \ArrowLine(-16,132)(0,116)

 \Text(-4,123)[lb]{$p_1$}
 \Text(-4,3)[lb]{$p_2$}
 \Text(82,123)[lb]{$p_4$}
 \Text(82,3 )[lb]{$p_3$}

\Text(98,63)[lb]{$q+p_3$}
\ArrowLine(94,55)(94,77)
\Text(-43,63)[lb]{$q-p_2$}
\ArrowLine(-5,77)(-5,55)

\end{picture}
\]
\vspace*{-7mm}
\caption[Example of a $J^d_{WA}(Q^2,T^2;b,\bar{t},d,\bar{u},W)$ function.]
        {Example of a $J^d_{WA}(Q^2,T^2;b,\bar{t},d,\bar{u},W)$ function.}
\label{WAbox}
\end{figure}

To calculate the weak part of the RC we introduce the set of auxiliary functions
$J^{d,c}_{WA,AW}$ related to ``direct'' and ``cross'' $WA$ and $AW$ box
diagrams of the kind shown in Fig.~\ref{WAbox}.
They are deeply connected to the procedure of separation of infra-red and mass singularities from Passa\-rino--Veltman
$D_0$ functions in terms of simplest objects --- the $C_0$ functions. The eventually ``subtracted'' auxiliary
functions $J_{sub}$ do not contain any singularities and are expressed as linear combinations of dilogarithms,
see~\cite{BKR}.
By introducing these functions we prove, first of all, that the EW part of the one-loop correction is free from
mass singularities and, moreover, receive a good profit in the stability and speed of numerical
calculations. Furthermore, the explicit expressions for these functions are used for the study of ``on-shell-W-mass''
singularities, introduced and discussed in Ref.~\cite{Wackeroth:1996hz}.

The letters $u,\, d,\, \ldots$ in the figure caption denote particle masses.
The ordering of masses in the argument of $J_{WA}^d$ into two pairs of heavy
(b,t) and light (d,u) quarks is such that the first mass in each pair corresponds to the fermion coupled
to the photon, thereby leading to the appearance of a potentially mass singular logarithmic contribution.

The basic definition of the function $J^d_{WA}$ reads:
\bqa
i\pi^2J^d_{WA}(Q^2,T^2;b,\bar{t},d,\bar{u},W)=\mu^{4-n}\int d^nq\frac{2 q\cdot p_1}{d_0d_1d_2d_3}\,,
\nonumber
\eqa
where
\bqa
d_0&=&(q-p_1-p_2)^2+\mw^2\,,\quad d_1\;=\;(q-p_2)^2+\mbtll\,,
\nll
d_2&=&q^2\,,\quad d_3\;=\;(q+p_3)^2+\mdnll\,.
\eqa

For $t$ and $\bar{t}$ decays one finds eight functions, four direct and four crossed ones.
The four direct ones come in two pairs:

\vspace*{-3mm}
\begin{figure}[!h]
\[
\begin{array}{ccc}
\begin{picture}(132,132)(-30,0)

\ArrowLine(0,110)(-22,132)
\Vertex(0,110){2.5}
\Text(-15,130)[lb]{$\bar b$}
                        \ArrowLine(66,132)(44,110)
                        \Photon(44,110)(0,110){3}{10}
                        \Vertex(44,110){2.5}
                        \Text(20,117)[lb]{$\gamma$}
                        \Text(55,130)[lb]{$\bar{d}$}
\ArrowLine(0,66)(0,110)
\ArrowLine(-22,44)(0,66)
\Vertex(0,66){2.5}
\Text(-15,40)[lb]{$t$}
                        \ArrowLine(44,110)(44,66)
                        \Vertex(44,66){2.5}
                        \Photon(0,66)(44,66){3}{5}
                        \Text(20,50)[lb]{$W$}
                        \ArrowLine(44,66)(66,44)
                        \Text(55,40)[lb]{$u$}
\Text(-30,20)[lb]{$ \small J^d_{AW}(..,\bar{b},t,\bar{d},u,W)\qquad$}
\Text(75,80)[lb]{$ =$}
\end{picture}
&
\begin{picture}(132,132)(0,0)
\ArrowLine(0,110)(-22,132)
\Vertex(0,110){2.5}
\Text(-15,130)[lb]{$\bar t$}
                        \ArrowLine(66,132)(44,110)
                        \Text(20,117)[lb]{$W$}
                        \Photon(44,110)(0,110){3}{5}
                        \Vertex(44,110){2.5}
                        \Text(55,130)[lb]{$\bar u$}
\ArrowLine(0,66)(0,110)
\ArrowLine(-22,44)(0,66)
\Vertex(0,66){2.5}
\Text(-15,40)[lb]{$b$}
                        \ArrowLine(44,110)(44,66)
                        \Vertex(44,66){2.5}
                        \Photon(0,66)(44,66){3}{10}
                        \Text(20,50)[lb]{$\gamma$}
                        \ArrowLine(44,66)(66,44)
                        \Text(55,40)[lb]{$d$}

\Text(-30,20)[lb]{$ \small J^d_{WA}(..,b,\bar t,d,\bar u,W)$}
\end{picture}
\end{array}
\]
\vspace*{-12mm}
\caption[First pair of the direct $J^d_{AW}$ and $J^d_{WA}$ functions.]
        {First pair of the direct $J^d_{AW}$ and $J^d_{WA}$ functions.}
\label{seriesjdirAW1}
\end{figure}
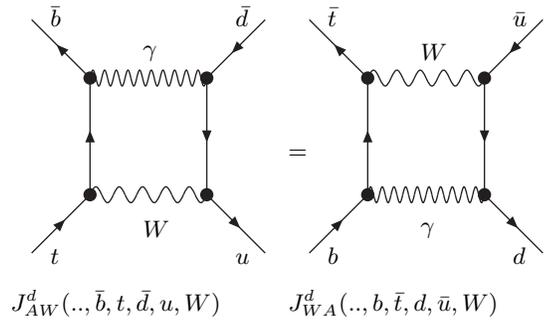

\vspace*{-5mm}
\begin{figure}[!h]
\[
\begin{array}{ccc}
\begin{picture}(132,132)(-40,0)
\ArrowLine(0,110)(-22,132)
\Vertex(0,110){2.5}
\Text(-15,130)[lb]{$\bar b$}
                        \ArrowLine(66,132)(44,110)
                        \Photon(44,110)(0,110){3}{5}
                        \Vertex(44,110){2.5}
                        \Text(20,117)[lb]{$W$}
                        \Text(55,130)[lb]{$\bar{d}$}
\ArrowLine(0,66)(0,110)
\ArrowLine(-22,44)(0,66)
\Vertex(0,66){2.5}
\Text(-15,40)[lb]{$t$}
                        \ArrowLine(44,110)(44,66)
                        \Vertex(44,66){2.5}
                        \Photon(0,66)(44,66){3}{10}
                        \Text(20,50)[lb]{$\gamma$}
                        \ArrowLine(44,66)(66,44)
                        \Text(55,40)[lb]{$u$}
\Text(-30,20)[lb]{$ \tiny J^d_{WA}(..,t,\bar{b},u,\bar{d},W)$}
\Text(75,80)[lb]{$ =$}
\end{picture}
&
\begin{picture}(132,132)(-10,0)
\ArrowLine(0,110)(-22,132)
\Vertex(0,110){2.5}
\Text(-15,130)[lb]{$\bar t$}
                        \ArrowLine(66,132)(44,110)
                        \Photon(44,110)(0,110){3}{10}
                        \Vertex(44,110){2.5}
                        \Text(20,117)[lb]{$\gamma$}
                        \Text(55,130)[lb]{$\bar{u}$}
\ArrowLine(0,66)(0,110)
\ArrowLine(-22,44)(0,66)
\Vertex(0,66){2.5}
\Text(-15,40)[lb]{$b$}
                        \ArrowLine(44,110)(44,66)
                        \Vertex(44,66){2.5}
                        \Photon(0,66)(44,66){3}{5}
                        \Text(20,50)[lb]{$W$}
                        \ArrowLine(44,66)(66,44)
                        \Text(55,40)[lb]{$d$}
\Text(-30,20)[lb]{$ \small J^d_{AW}(..,\bar{t},b,\bar{u},d,W)$}
\end{picture}
\end{array}
\]
\vspace*{-12mm}
\caption[Second pair of direct $J^d_{WA}$ and $J^d_{AW}$ functions.]
        {Second pair of direct $J^d_{WA}$ and $J^d_{AW}$ functions.}
\label{seriesjdirAW2}
\end{figure}
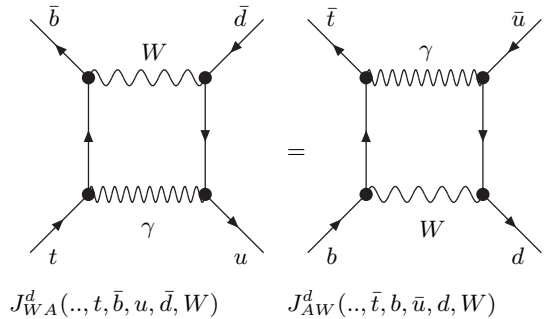
\vspace*{-5mm}

The crossed functions 
may be obtained by a simple permutation of their arguments.

There are four symmetry relations between direct $J^d_{AW,WA}$ and cross $J^c_{AW,WA}$ functions:
\bqa
J^d_{AW}(\Qs,\Ts;\bar{b},t,\bar{d},u,W) = J^d_{WA}(\Qs,\Ts;b,\bar{t},d,\bar{u},W), \nll
J^d_{WA}(\Qs,\Ts;t,\bar{b},u,\bar{d},W) = J^d_{AW}(\Qs,\Ts;\bar{t},b,\bar{u},d,W), \nll
J^c_{AW}(\Qs,\Us;\bar{b},t,u,\bar{d},W) = J^c_{WA}(\Qs,\Us;b,\bar{t},\bar{u},d,W), \nll
J^c_{WA}(\Qs,\Us;t,\bar{b},\bar{d},u,W) = J^c_{AW}(\Qs,\Us;\bar{t},b,d,\bar{u},W). \nonumber
\eqa

So, only four functions are independent. Moreover, as seen from the previous
relations, 
the indices content of the $J^d_{AW}\dots$ functions (retained for better understanding of their origin from
corresponding Feynman diagrams) is uniquely determined by their arguments. Therefore, these indices
may be dropped in the subsequent presentation of the material.
Also the particle names will be changed to particle masses in the arguments of these functions.

\subsubsection{Steps to calculate $J$ functions}

$\bullet$ \underline{step: relations for $J$}

Using the standard Passarino--Veltman reduction it is possible to establish relations
(exact in masses) between infra-red divergent functions
(from here and below, we use the usual notation for particle masses) \\
$D_0( - \mbtll, - \mtpll, - \mupll, - \mdnll,\Qs,\Ts;0,\mbt,\mw,\mdn)$, \\
$C_0( - \mdnll, - \mbtll,\Ts;\mdn,0,\mbt)$\\
and infra-red finite, but mass-singular functions:\\
$J(\Qs,\Ts;\mbt,\mtp,\mdn,\mup,\mw)$ \\ and
$C_0( - \mupll, - \mdnll,\Qs;\mw,\mdn,0)$.

For direct functions these relations are:
\bqa
&& J(\Qs,\Ts;\mbt,\mtp,\mdn,\mup,\mw) = (\mwll+\Qs)\times  \nll
&& D_0( - \mbtll, - \mtpll, - \mupll, - \mdnll,\Qs,\Ts;0,\mbt,\mw,\mdn)
\nll
&& + C_0( - \mupll, - \mdnll,\Qs;\mw,\mdn,0)    \nll
&& - C_0( - \mdnll, - \mbtll,\Ts;\mdn,0,\mbt)\,,\nll
&& J(\Qs,\Ts;\mtp,\mbt,\mup,\mdn,\mw) = (\mwll+\Qs) \times
\nll
&& D_0( - \mbtll, - \mtpll, - \mupll, - \mdnll,\Qs,\Ts;\mw,\mtp,0,\mup)
\nll
&& + C_0( - \mupll, - \mdnll,\Qs;0,\mup,\mw) \nll
&& - C_0( - \mtpll, - \mupll,\Ts;\mtp,0,\mup)\,.
\eqa
For the crossed functions we perform the appropriate permutations of the arguments of these functions.

Then we calculate the functions $J$ exactly in masses in terms of dilogarithms.
Finally, we take the limit $\mup,\mdn \rightarrow 0$, neglecting light quark masses
everywhere but mass singular logarithms.
These two steps represent rather complicated precedures, which will be described elsewhere~\cite{BKK:2009prep}.

$\bullet$ \underline{step: $J_{sub}$}

The mass singularities in arguments of the logarithms may be compensated
by combination with one more $C_0$ function:
\bqa
\label{Jsub}
&& J_{sub}(\Qs,\Ps;\mbt,\mtp,\mw)  =       \\
&& J(\Qs,\Ps;\mbt,\mtp,\mdn,\mup,\mw)      \nll
&& -\left(1+\frac{\Qs}{\mbtll+\Ps}\right) C_0(-\mupll,-\mdnll,\Qs;\mw,\mdn,0)\,,
\nll
&& J_{sub}(\Qs,\Ps;\mtp,\mbt,\mw)  =        \nll
&& J(\Qs,\Ps;\mtp,\mbt,\mup,\mdn,\mw)      \nll
&& -\left(1+\frac{\Qs}{\mtpll+\Ps}\right) C_0(-\mdnll,-\mupll,\Qs;\mw,\mup,0)\,.
\nonumber
\eqa
where $\Ps=\Ts$ or $\Ps=\Us$.
The two mass-singular $C_0$ functions appearing in Eq.~(\ref{Jsub}) cancel in the total
expression for the EW correction which proves the absence in it of logarithmic mass
singularities (not KLN theorem!).

$\bullet$ \underline{step: $J_{subsub}$}

If we want to neglect the $\mbt$-mass, we should perform a second subtraction of
a mass singular $C_0$ function $C_0(-\mtpll,-\mbtll,\Qs,\mw,\mbt,0)$ that
appears in the limit  $\mbt=0$.

Note that only one of $J_{sub}$ contains an $\mbt$ mass singularity.
\bqa
&& J_{subsub}(\Qs,\Ps;\mbt,\mtp,\mw) = \nll
&& J_{sub}(\Qs,\Ps;\mbt,\mtp,\mw)
\nll
&&  -\frac{\Ps}{\Qs+\mtpll} C_0(-\mtpll,-\mbtll,\Qs;\mw,\mbt,0).
\eqa

Since we do not want to consider the limit $\mtp=0$, we simply rename the second function:
\bqa
&& J_{subsub}(\Qs,\Ps;\mtp,\mbt,\mw) = \nll
&& J_{sub}(\Qs,\Ps;\mtp,\mbt,\mw).
\eqa

Again, the $\mbt$ mass singular $C_0$ function \\
$C_0(-\mtpll,-\mbtll,\Qs;\mw,\mbt,0)$
cancels in the total EW correction.


\subsubsection{Treatment of on-shell-W-mass singularities\label{OMSWsing}}
In the course of calculations of the ${\cal{O}}{(\alpha)}$ EWRC one encounters
{\em on-shell} singularities which appear in the form of $\ln(s-\mw^2+i\epsilon)$.
We follow Ref.~\cite{Wackeroth:1996hz} where it was shown that they can be regularized by the $W$ width:
\bqa
\ln(s-\mw^2+i\epsilon) \to \ln(s-\mw^2+i\mw\gw).
\label{onshellreg}
\eqa
Note that the replacement $\mw^2-i\epsilon\to\mw^2-i\mw\gw$ should be done only in the argument of logarithms
which diverge at the resonance $s=\mw^2$. In this connection we derived for all $J_{subsub}$ functions
such a representation in which these divergent logarithms appear only once and $\gw$ propagates only in it.
Everywhere else we retain $\mw^2-i\epsilon$. The explicit formulae for $J_{subsub}$ functions will be
presented elsewhere Ref.~\cite{BKK:2009prep}.

We also meet the on-shell singular $C_0$ and $B_0$ functions. They correspond to non-abelian $Wff'$ vertex
functions with a virtual photon coupled to one of the fermions of mass $m$ and to a $W$ boson
and to the $W$ boson self-energy diagram, respectively. We give explicit expressions for both functions:
\bqa
&&C_0(0,-m^2,-s;\mw,m,0)=\frac{1}{m^2-s}\biggl[\ln\left(\frac{s}{m^2}\right)
\nll &&\times
       \ln\left(-\frac{s-\mwll+i\mw\gw}{m^2-\mwll}\right)
      -\frac{1}{2}\ln^2\left(\frac{s}{m^2}\right)
\nll &&
      -\Litwo\left(\frac{m^2(-s+\mwll-i\epsilon)}{s(\mwll-m^2)}\right)
      -\Litwo\left(\frac{s}{m^2-i\epsilon}\right)
\nll &&
      +\Litwo\left(\frac{-s+\mwll-i\epsilon}{\mwll-m^2}\right)
      +\Litwo(1)\biggr],
\eqa
where the first ``0'' stands for a fermion whose mass may be ignored (neutrino or $b$-quark); and
\bqa
&&B_0^F(-s,\mu^2;\mw,0)=2-\ln\left(\frac{\mwll}{\mu^2}\right)-\left(1-\frac{\mwll}{s}\right)
\nll&&\times
          \ln\left(-\frac{s-\mwll+i\mw\gw}{\mwll}\right).
\eqa

\section{Cascade approximations \label{Cascade}}
\subsection{The usual narrow width cascade\label{Cascadenarrow}}
In this approach we create a narrow width cascade using one-loop $t\to Wb$ and $W\to l \nu$
formulae, i.e.
\bq
\Gamma_{t\to bl\nu}=
\frac{\Gamma^{\rm 1loop}_{t\to Wb}\Gamma^{\rm 1loop}_{W\to l\nu}}{\gw}\,.
\label{cascad}
\eq
At one-loop, it is more consistent to use instead its ``linearized'' version
\bq
\Gamma_{t\to bl\nu}=
\frac{\Gamma^{\rm Born}_{t\to Wb}   \Gamma^{\rm Born}_{W\to l\nu}}{\gw}
\left(1+\delta^{\rm 1loop}_{t\to Wb}+\delta^{\rm 1loop}_{W\to l\nu}\right),
\label{cascadli}
\eq
where~~$\delta^{\rm 1loop}=\Gamma^{\rm 1loop}/\Gamma^{\rm Born}-1\,.$

\subsection{Cascade with complex $W$ mass\label{Cascadecmplx}}
Another approach to the one-loop cascade approximation uses the same Eq.~(\ref{cascad})
but with a complex $W$ mass,
\bq
\tmW^2=\mw^2-i\mw\gw\,,
\label{mwshift}
\eq
in all $W$ boson propagators in the diagrams with radiation of real or virtual photons.
This trick regula\-rizes the corresponding infrared divergences. The modified Passarino--Veltman
functions are listed below in this section and the results of new calculations are discussed
in section~\ref{Numerics}.

This modification affects all infrared divergent loop and brem\-sstrahlung
diag\-rams where a photon is coupled to the $W$ boson: they all become infrared finite.

The modification of the calculation is trivial for squares and interferences of the
corresponding brems\-strahlung diagrams, which after the replacement
$\mw^2\to\tmW^2$ may be treated like infrared stable hard photon contributions.
For loop diag\-rams one should replace infrared divergent PV functions in expressions
regularized by $\gw$. They are listed below.

\subsubsection{Analytic expression for modified PV functions\label{pvfs}}
The infrared divergent derivative
$B^\prime_0(-\mw^2;0,\tmW)=[dB_0(p^2;0,\tmW)/dp^2]_{|p^2=-{M_{\sss W}^2}}$
of the $B_{0}$ function, which arises from a counterterm
related to the $W$ boson self-energy diagram, becomes:
\bqa
B^\prime_0(-\mw^2;0,\tmW)
=\frac{1}{\mw^2}\left[1+\ln \left(\frac{\tmW^2-\mw^2}{\mw^2}\right)\right].\quad\;\;
\eqa
There is only one generic $C_0$ 3-point function with a photon coupled to
the $W$ boson and a fermion with mass $m_2$; $m_1$ is the mass of the other
fermion:
\bqa
&&C_0(-m^2_1,-m^2_2,-\mwll;{\tmW},m_2,0)=
\nll && \frac{1}{S_{l}} \Biggl\{
          \Biggl[-\ln\left(\frac{\tmW^2-\mwll}{m^2_2}\right)l\left(y_{l_1}\right)
\nll &&
          +\frac{1}{2}l^2\left(y_{l_1}\right)
          +\ln\left(1-\frac{y_{l_1}}{y_{l_2}}\right)l\left(y_{l_1}\right)
\nll &&
          -\Litwo\left( \frac{1-y_{l_1}}{y_{l_2}-y_{l_1}}\right)
          +\Litwo\left(-\frac{  y_{l_1}}{y_{l_2}-y_{l_1}}\right)
\nll &&
          -\Litwo\left( \frac{1}{y_{l_1}}\right)
          \Biggr]
        - \Biggl[ y_{l_1} \leftrightarrow  y_{l_2} \Biggr]
                        \Biggr\}.
\label{c02cmplx}
\eqa
Here
\bq
l(y)=\ln\left(1-\frac{1}{y}\right),
\eq
and
\bqa
y_{l_1}&=&\frac{m_1^2+m_2^2-\mwll+i\epsilon+S_{l}}{2 m_1^2}\,,
\nll
y_{l_2}&=&\frac{m_1^2+m_2^2-\mwll+i\epsilon-S_{l}}{2 m_1^2}\,,
\nll
S_{l} &=&\sqrt{(m_1^2+m_2^2-\mwll+i\epsilon)^2-4 m_1^2 m_2^2}\,.
\label{c01cmplx}
\eqa
Its limit where the radiating mass $m_2$ is arbitrary and the other fermion mass
is zero is much more compact:
\bqa
&&C_0(0,-m^2_2,Q^2;{\tmW},m_2,0)=
\frac{1}{M_{\sss W}^2-m_2^2} \Biggl[
\nll &&
           -\ln\left(\frac{{\tmW}^2-\mwll}{m_2^2}\right)l\left(y_{l}\right)
      + \frac{1}{2}l^2\left(y_{l}\right)     -\Litwo\left(\frac{1}{y_{l}}\right)
                                   \Biggr],
\nll
\eqa
where
\bq
y_l = \frac{m_2^2}{m_2^2-\mwll+i\epsilon}\,.
\eq
Finally in the limit $m_2\to 0$, Eq.~(\ref{c02cmplx}) simplifies to
\bqa
\label{c02li}
&&C_0(-m^2_1,-m^2_2,-\mwll;{\tmW},m_2,0)=\\
&& \frac{1}{m^2_1-\mwll} \biggl[-\ln\left(\frac{m^2_2}{m^2_1}\right)\ln\left(\frac{\tmW^2-\mwll}{m^2_1-\mwll}\right)
\nll &&+\ln\left(\frac{\tmW^2-\mwll}{m^2_1}\right)\bigl[2\ln\left(-y_l\right)-\ln\left(1-y_l\right)\bigr]
\nll &&+\frac{1}{2}\ln^2\left(1-y_l\right)-2\ln^2\left(-y_l\right)-\Litwo\left(y_l\right)-2\Litwo(1) \biggr],
\nonumber
\eqa
where
\bq
y_l=\frac{m^2_1-\mwll+i\epsilon}{m^2_1}\,.
\eq
In Eq.(\ref{c02li}) the mass singular term is separated out explicitly. This expression
is especially convenient if one wants to control mass singularities.

\subsection{Pole approximation\label{CascadeBW}}
Here we present the cascade pole approximation with the aid of the two one-loop building blocks as illustrated
in Fig.~\ref{BW}. This gives a schematic representation of a convolution of a Breit--Wigner distribution for
a virtual $W$ boson with two pairs of building blocks: one at one-loop level (big blob) and the second one
at tree level, and vice versa.

First, define the one-loop corrected decay width for two decays
{\em off the $W$ mass shell at some} $\mws$; 
\bqa
\Gamma^{\mathrm{1loop}}_{t\to Wb}(\mws)&
=&\Gamma^{\mathrm{Born}}_{t\to Wb}(\mws)\left[1+\delta^{\mathrm{weak}}_{t\to Wb}(\mwll)\right]
\qquad
\nll
&&+\Gamma^{\mathrm{virtsoft}}_{t\to Wb}(\mws)+\Gamma^{\mathrm{hard}}_{t\to Wb}(\mws),
\eqa
and a similar representation for the $W\to l\nu$ decay.

Note that $\delta^{\mathrm{weak}}$ is frozen at $\mwll$.
This trick ensures an approximate gauge invariance for CC processes
(for NC processes it would ensure exact gauge invariance).

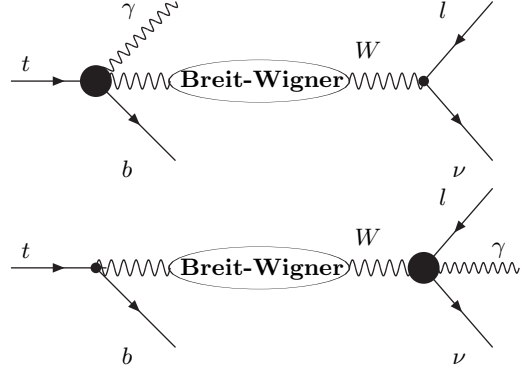
\begin{figure}[!t]
\vspace*{-12mm}
\[
\begin{array}{cc}
\begin{picture}(132,132)(0,0)
 \ArrowLine(8,66)(44,66)
 \Vertex(40,66){6}
 \Text(12, 70)[lb]{$t$}
\ArrowLine(40,66)(70,36)
\Photon(40,66)(68,66){3}{6}
\Photon(40,66)(70,96){2}{10}
 \Text(50,90)[lb]{$\gamma$}
 \Text(50,30)[lb]{$b$}
\end{picture}
&
\begin{picture}(132,132)(15,0)
 \Photon(16,66)(44,66){3}{6}
 \Vertex(44,66){2}
 \Text(18, 75)[lb]{$W$}
\ArrowLine(70,96)(44,66)
\ArrowLine(44,66)(70,36)
 \Text(50,90)[lb]{$l$}
 \Text(55,30)[lb]{$\nu$}
 \Text(-48,62)[lb]{\bf \small Breit-Wigner}
 \Oval(-18,66)(34,8)(-90)
\end{picture}
\end{array}
\]
\\
\vspace*{-35mm}
\[\begin{array}{cc}
\begin{picture}(132,132)(0,0)
 \ArrowLine(8,66)(44,66)
 \Vertex(40,66){2}
 \Text(12, 70)[lb]{$t$}
\ArrowLine(40,66)(70,36)
\Photon(40,66)(68,66){3}{6}
 \Text(50,30)[lb]{$b$}
\end{picture}
&
\begin{picture}(132,132)(15,0)
 \Photon(16,66)(44,66){3}{6}
 \Vertex(44,66){6}
 \Text(18, 75)[lb]{$W$}
\ArrowLine(70,96)(44,66)
\ArrowLine(44,66)(70,36)
 \Text(50,90)[lb]{$l$}
 \Text(55,30)[lb]{$\nu$}
 \Text(-48,62)[lb]{\bf \small Breit-Wigner}
 \Oval(-18,66)(34,8)(-90)
\Photon(44,66)(80,66){2}{10}
 \Text(70,70)[lb]{$\gamma$}
\end{picture}
\end{array}
\]
\vspace*{-15mm}
\caption[$t\to bf_{1}f_{1}$ decay \label{BW}]
        {$t\to bf_{1}f_{1}$ decay \label{BW}.}
\vspace*{-5mm}
\end{figure}

The one-loop $\Gamma^{\mathrm{1loop}}_{t\to bl\nu}$ is given by the following convolution integral:
\bqa
\Gamma^{\mathrm{1loop}}_{t\to bl\nu}&=&\frac{1}{k}\int^u_ld\mws\biggl[
 \Gamma^{\mathrm{1loop}}_{t\to Wb}(\mws)\Gamma^{\mathrm{Born}}_{W\to l\nu}(\mws)
\nll
&&+\Gamma^{\mathrm{1loop}}_{W\to l\nu}(\mws)\Gamma^{\mathrm{Born}}_{t\to Wb}(\mws)
\nll
&&-\Gamma^{\mathrm{Born}}_{t\to Wb}(\mws)\Gamma^{\mathrm{Born}}_{W\to l\nu}(\mws)
\biggr]
\nll
&&\times
\frac{\mw}{(\mws-\mwll)^2+\mwll\gw^2}\,,
\eqa
where $k$ is given by the normalization of the Breit--Wigner distribution and
$u$ and $l$ are the broadest limits allowed by the decay kinematics:
\bqa
 k&=&\atan(k_{min})+\atan(k_{max})\,,
\nll
 u&=&\mwll+k_{max}\mw\gw\,,
\nll
 l&=&\mwll-k_{min}\mw\gw\,,
\nll
 k_{min}&=&\frac{\mwll-m^2_l}{\mw\gw}\,,
\nll
 k_{max}&=&\frac{(\mtp-\mbt)^2-\mwll}{\mw\gw}\,,
\eqa
where $m_l$ is the charged lepton mass.

This {\em finite width approximation}, as one may call it, allows a fully differential realization,
and hence also MC generation.

\section{Numerical results\label{Numerics}}
We present all numbers, computed with the standard \sanc {\tt INPUT}, PDG(2006)~\cite{PDG2006}:
\bqa \nonumber
\begin{array}[b]{lcllcllcllcl}
&G_{\sss F} & = & 1.16637\cdot 10^{-5}\GeV^{-2}, &\alpha(0) &=& 1/137.03599911, \\
&\mw & = & 80.403 \GeV, & \gw & = & 2.141\GeV, \\
&\mz & = & 91.1876\GeV, & \mh & = & 120\GeV,   \\
&m_e & = & 0.51099892\MeV,& m_u & = & 62\;\MeV,\\
&m_d & = &  83\;\MeV, & m_{\tau} &=&1.77699\GeV,\\
&m_c & = & 1.5\;\GeV, & m_s & = & 215\;\MeV,   \\
&m_b & = & 4.7\;\GeV, & m_t & = & 174.2\;\GeV, \\
&m_{\mu}&=&0.105658369\GeV, & \alpha_{s}&=&0.107.\\
\end{array}
\label{sanc2006}
\eqa

First, we illustrate the dependence of the complete one-loop EW results on $\mbt$ for two
decay channels and two ways of calculations, with and without taking account of $\gw$
to regularize on-shell $W$ boson singularities as discussed in section~\ref{OMSWsing}.
Table~\ref{tab3} shows QCD NLO results, where the account of $\gw$ is irrelevant since
the gluons are not coupled to the $W$ boson.

\begin{table}[!h]
\begin{center}
\begin{tabular}{|c||c|c|c|c|}
\hline
{$\mbt,$}&
\multicolumn{2}{|c|} {$t\to bl^+\bar{\nu_l}$}&
\multicolumn{2}{|c|} {$t\to bu\bar{d}$}\\
\cline{2-5}
{GeV}
&$\Gamma^{\mathrm{1l}},\,$MeV&$\delta,\%$
&$\Gamma^{\mathrm{1l}},\,$MeV&$\delta,\%$\\
\hline
{4.7}      & 159.877(3) & 6.953(2) & 480.341(6) & 7.111(1)\\
\hline
{1.0}      & 159.872(3) & 6.949(2) & 480.339(6) & 7.111(1)\\
\hline
{0.1}      & 159.871(3) & 6.949(2) & 480.337(6) & 7.110(1)\\
\hline
\end{tabular}
\vspace*{2mm}
\caption{One-loop decay widths $\Gamma^{\mathrm{1l}}$
and percentage of the EWRC for complete calculations
in $\alpha(0)$-scheme as a function of the $\mbt$ mass and with $\gw$
kept only in on-shell $W$ boson singular terms.\label{tab1}}
\end{center}
\end{table}

As seen from Tables~\ref{tab1}-\ref{tab3}, EW and QCD corrections have the opposite sign and
QCD corrections are relatively larger. The $\mbt$ dependence is barely visible in
$\Gamma^{\mathrm{1l}}$ and consistent with no-dependence in $\delta$ within the statistical
errors. This allows us to simplify the analysis and to present all the subsequent results at
a small $\mbt$ using simplified formulae for weak one-loop contributions for $\mbt=0$.
The QED/QCD contributions contain $\ln(\mbt)$ in different parts but they
cancel in the sum in accordance with the KLN theorem.
Tables~\ref{tab1}-\ref{tab3} demonstrate the validity of the KLN theorem.

\begin{table}[!t]
\begin{center}
\begin{tabular}{|c||c|c|c|c|}
\hline
{$\mbt$,}&
\multicolumn{2}{|c|} {$t\to bl^+\bar{\nu_l}$}&
\multicolumn{2}{|c|} {$t\to bu\bar{d}$}\\
\cline{2-5}
{GeV}
&$\Gamma^{\mathrm{1l}},\,$MeV&$\delta,\%$
&$\Gamma^{\mathrm{1l}},\,$MeV&$\delta,\%$\\
\hline
{4.7}      & 159.943(3) & 6.997(2) & 480.661(6) & 7.183(1)\\
\hline
{1.0}      & 159.938(3) & 6.993(2) & 480.658(6) & 7.182(1)\\
\hline
{0.1}      & 159.937(3) & 6.993(2) & 480.656(6) & 7.182(1)\\
\hline
\end{tabular}
\vspace*{2mm}
\caption{One-loop decay widths $\Gamma^{\mathrm{1l}}$
and percentage of the EWRC for complete calculations in $\alpha(0)$-scheme as a function of the
$\mbt$ mass and without regularization of on-shell $W$ boson singularities.\label{tab2}}
\end{center}
\end{table}

\begin{table}[!h]
\begin{center}
\begin{tabular}{|c||c|c|c|c|}
\hline
{$\mbt$,}&
\multicolumn{2}{|c|} {$t\to bl^+\bar{\nu_l}$}&
\multicolumn{2}{|c|} {$t\to bu\bar{d}$}\\
\cline{2-5}
{GeV}
&$\Gamma^{\mathrm{1l}},\,$MeV&$\delta,\%$
&$\Gamma^{\mathrm{1l}},\,$MeV&$\delta,\%$\\
\hline
{4.7}      & 136.73(2)  & -8.53(1) & 358.72(28) & -20.01(6) \\
\hline
{1.0}      & 136.70(4)  & -8.55(2) & 358.04(31) & -20.16(7) \\
\hline
{0.1}      & 136.69(6)  & -8.56(4) & 358.87(35) & -19.98(8) \\
\hline
\end{tabular}
\vspace*{2mm}
\caption{One-loop decay widths $\Gamma^{\mathrm{1l}}$
and percentage of the QCD correction for complete calculations
in $\alpha(0)$-scheme as a function of the $\mbt$ mass.\label{tab3}}
\end{center}
\end{table}

For definiteness, the numbers presented in the following Tables, after Table~\ref{tab3},
are computed for $\mbt = 1$GeV, since even at $\mbt = 4.7$GeV the numbers are practically
the same as at $\mbt = 0.1$GeV.

In Table~\ref{tab4} we illustrate the $\gw$ dependence of EWRC to the two channels under consideration,
irrelevant for QCD NLO corrections.
\begin{table}[!h]
\begin{center}
\begin{tabular}{|c||c|c|c|c|}
\hline
{$\frac{\ds\gw}{\ds N}$}&
\multicolumn{2}{|c|} {$t\to bl^+\bar{\nu_l}$}&
\multicolumn{2}{|c|} {$t\to bu\bar{d}$}\\
\hline
{$N$}
&$\Gamma^{\mathrm{1l}},\,$MeV&$\delta,\%$
&$\Gamma^{\mathrm{1l}},\,$MeV&$\delta,\%$\\
\hline
$1$        & 159.872(3) & 6.949(2) & 480.339(6) & 7.111(1)\\
\hline
$10$       & 159.943(3) & 6.997(2) & 480.638(6) & 7.177(1)\\
\hline
$10^2$     & 159.938(3) & 6.994(2) & 480.656(6) & 7.182(1)\\
\hline
$10^3$     & 159.938(3) & 6.993(2) & 480.658(6) & 7.182(1)\\
\hline
\hline
{$\infty$} & 159.938(3) & 6.993(2) & 480.658(6) & 7.182(1)\\
\hline
\end{tabular}
\vspace*{2mm}
\caption{One-loop decay widths and percentage of the EWRC for complete calculations
in $\alpha(0)$-scheme as a function of $\gw$.\label{tab4}}
\end{center}
\end{table}

This Table illustrates the perfect convergence with lowering $\gw$ and consistency
of numbers for $\gw/10^2$ with results computed with zero width in arguments of functions with on-shell $W$ mass
singularities, see section~\ref{OMSWsing}.

Now turn to the study of narrow width cascade approaches, see section~\ref{Cascade}.
All numbers are presented in the $\alpha(0)$-scheme for definiteness.
Here we limit ourselves to EWRC, because of the vanishing of $gW$ boxes in the QCD case.
Comparison of complete and cascade approaches shows in particular the importance of EW boxes
which are absent in the cascade approach. Two $\delta$'s are shown corresponding to Eq.~(\ref{cascad}),
factorized version, and Eq.~(\ref{cascadli}), linearized version.

\begin{table}[!ht]
\begin{center}
\begin{tabular}{|c||c|c|c|c|c|}
\hline
&{$t\to Wb$}&{$W\to e\nu$}&{$t\to be\nu$ }\\[-4mm]
&           &             &{ cascade }    \\
\hline
$\Gamma^{\mathrm{Born}}$, MeV & 1480.0 & 219.70 & 151.87 \\
$\Gamma^{\mathrm{1l}}$,   MeV & 1546.6 & 225.28 & 162.73 \\
$\delta, \%$                 & 4.495  & 2.538  & 7.155  \\
$\delta_{\rm lin},\%$         &        &        & 7.033  \\
\hline
\end{tabular}
\end{center}
\vspace*{2mm}
\caption{Born, one-loop decay widths and percentage of the correction in narrow width cascade
approximation, $\alpha(0)$-scheme.\label{tab5}}
\end{table}

Table~\ref{tab5} shows rather good agreement of complete and narrow width cascade calculations
for inclusive quantities. The linearized version agrees better.
This is natural, since the complete calculations in \sanc are linearized by default.

Next Table~\ref{tab6} shows the results of the cascade approach with complex $W$ mass, see
section~\ref{Cascadecmplx}.

\begin{table}[!ht]
\begin{tabular}{|c||c|c|c|c|c|c|}
\hline
$\frac{\ds\gw}{\ds N} $
 &
\multicolumn{2}{|c|} {$t\to Wb$}&
\multicolumn{2}{|c|} {$W\to e\nu$}&
\multicolumn{2}{|c|} {$t\to bl\nu$}\\[-4mm]
 &
\multicolumn{2}{|c|}  { }      &
\multicolumn{2}{|c|}  { }      &
\multicolumn{2}{|c|} { cascade} \\
\hline
N
&\hsm$\Gamma_{t\to Wb}$,\hsm
&$\delta$,
&\hsm$\Gamma_{W\to e\nu}$,\hsm
&$\delta$,
&\hsm$\Gamma_{t\to bl\nu}$,\hsm
&$\delta$, \\[-3mm]
& MeV &\%& MeV &\%& MeV &\%\\
\hline
\hsm $ 1   $ \hsm & \hsm 1543.4 \hsm & 4.29 & \hsm 225.05 \hsm & 2.43 & \hsm 162.23 \hsm & 6.83 \\
\hsm $ 10  $ \hsm & \hsm 1543.0 \hsm & 4.26 & \hsm 224.79 \hsm & 2.32 & \hsm 162.00 \hsm & 6.68 \\
\hsm $ 10^2$ \hsm & \hsm 1543.0 \hsm & 4.26 & \hsm 224.77 \hsm & 2.31 & \hsm 161.99 \hsm & 6.67 \\
\hsm $ 10^3$ \hsm & \hsm 1543.0 \hsm & 4.26 & \hsm 224.77 \hsm & 2.31 & \hsm 161.99 \hsm & 6.67 \\
\hline
\end{tabular}
\vspace*{2mm}
\caption{One-loop decay widths and percentage
of the correction in cascade approximation with complex $W$ mass~.\label{tab6}}
\end{table}

There is again good convergence with decreasing $\gw$, however we see that
the agreement of this cascade version
with the complete one-loop calculation (see Table~\ref{tab1})
degrades with decreasing $W$ boson width.

\noindent
Finally, in Table~\ref{tab7} we present the results of calculations within
the finite width cascade approach in the pole approximation for the $t\to bl^+\bar{\nu_l}$ decay.

\begin{table}[!ht]
\begin{center}
\begin{tabular}{|c||c|c|c|}
\hline
{$\frac{\ds\gw}{\ds N}$}
&
\multicolumn{3}{|c|} {$t\to bl^+\bar{\nu_l}$}\\
\hline
{$ N $}
&$\Gamma^{\mathrm{Born}},\,$MeV&$\Gamma^{\mathrm{1l}},\,$MeV&$\delta,\%$\\
\hline
{$1$}    &153.244(1) &164.015(1) &7.029(1)\\
\hline
{$10$}   &152.007(1) &162.696(1) &7.032(1)\\
\hline
{$10^2$} &151.880(1) &162.561(1) &7.032(1)\\
\hline
{$10^3$} &151.868(1) &162.548(1) &7.032(1)\\
\hline
\end{tabular}
\vspace*{2mm}
\caption{Born, one-loop decay widths and percentage of the EWRC for the
pole approximation, $\alpha(0)$-scheme, as a function of $\gw$.\label{tab7}}
\end{center}
\end{table}

This is the main result of the study of the validity of resonance approaches and it deserves a detailed
discussion. By now we only note that there is convergence with decreasing $\gw$ and full consistency with
the narrow width cascade results. Since this approach is aimed at extending the cascade
approximation to the description of exclusive quantities, it is worth testing it for a simple
distribution, like $d\Gamma/ds$, where $s$ is the  invariant mass squared of the $l^+\bar{\nu_l}$ pair.

\section{Conclusions\label{Concl}}
We have described the work for the $t \to b f_{1} f'_{1}$ decays. We have
computed both QCD and EW total one-loop corrections within the SANC system for all decays.

We have discussed EW corrections in more detail as they are more complicated than QCD.
We have considered the problem of separating of the QED contribution from the complete EW correction.

 Auxiliary functions $J_{AW(WA)}^{d(c)}$ for these decays were introduced.
Then we have presented numerical results, obtained with the aid of a Monte Carlo
integrator.

 We study the $m_b$ dependence of EW and QCD corrections
showing the validity of the KLN theorem. We have also demonstrate the effect of taking account of the
$W$ width in the EW contribution.

A comprehensive research of using different cascade approximations in
numerical evaluations was done. The goal of this research was to check the
possibility of using building blocks calculated in SANC to construct construction the MC
tools for complicated actual processes.  We have studied the narrow width cascade,
cascade with complex $W$ mass approximations and cascade in the pole
approximation. The difference between cascade methods and complete
calculations shows the effect of EW boxes that are missed in the cascade
approaches. However it is relatively small and one can see rather good
agreement of cascade approaches with complete calculations. So, all these methods could be applied.

The most important here is the consideration of the case of pole
approximation, as it represents the differential realization of decay widths.
This allows the event generation within a cascade approach. However, the comparison
with the complete calculations at the level of differential event
distributions would be also required. That is the goal of a future work.

{\em Acknowledgements.} We are gratelul to A.~Arbuzov and L. Rumyantsev for discussions.

This work is partly supported by RFFI grant $N^{o}$07-02-00932-a; one of us
(V. Kolesnikov) thanks the Dynasty Foundation for support.

\providecommand{\href}[2]{#2}\end{document}